\documentclass{elsart}
\usepackage{epsfig}
\usepackage{amssymb}

\begin{document}

\begin{frontmatter}

\title{
Analyzing power of the ${\vec p}p \rightarrow pp{\pi}^0$ reaction at beam energy of 390 MeV}

 \author[label1,label7]{Y.Maeda\corauthref{cor1}},
 \author[label1]{M.Segawa}, 
 \author[label1]{H.P.Yoshida}, 
 \author[label2]{M.Nomachi},
 \author[label2]{Y.Shimbara},
 \author[label2]{ Y.Sugaya}, 
 \author[label3]{K.Yasuda},
 \author[label4]{K.Tamura}, 
 \author[label5]{T.Ishida},
 \author[label5]{T.Yagita},
 \author[label7,label6]{A.Kacharava}
 \address[label1]{
Research Center for Nuclear Physics, 
Osaka University, Ibaraki,Osaka 567-0047, Japan}
 \address[label7]{
Institut f{\"u}r Kernphysik, Forschungszentrum J{\"u}lich, 52425
 J{\"u}lich, Germany\thanksref{label8}}
\address[label2]{
Department of Physics, Osaka University, 
Toyonaka, Osaka 560-0043, Japan}
 \address[label3]{
The Wakasa Wan Energy Research Center, 
Fukui 914-0192, Japan}
 \address[label4]{
Physics Division, Fukui Medical University, 
Fukui 910-1193, Japan}
 \address[label5]{
Department of Physics, Kyushu University, Fukuoka
812-8581, Japan}
 \address[label6]{
Laboratory of Nuclear Problems, Joint Institute for Nuclear Research, Dubna, 141980, Russia
}
\corauth[cor1]{Corresponding author.
E-mail address: y.maeda@fz-juelich.de}
\thanks[label8]{Present address.}
\begin{abstract}
The analyzing power of ${\vec p}p\rightarrow pp{\pi}^0$ reaction has been measured at the beam energy of 390 MeV. 
The missing mass technique of final protons has been applied to identify the $\pi^0$ production event. 
The dependences of the analyzing power on the pion emission-angle and the relative momentum of the protons have been obtained. 
The angular dependence could be decomposed by the Legendre polynomial and the relative contribution of the $P_{21}$ to $P_{11}$ function is less than 20$\%$.
The $P$-state amplitude is found to be the dominant component of the $\pi$ production near the threshold. 
The momentum dependence of the analyzing power has been studied to obtain the information about the pion production mechanism.
It has been deduced that the pion production due to the long range interaction plays an important role in the momentum dependence of the $P$-state amplitude.
\end{abstract}

\end{frontmatter}

In the last few years,
both experimental and theoretical investigations for various $NN \rightarrow NN{\pi}$ reactions have been performed. 
The total cross section of the $pp \rightarrow pp{\pi}^0$ reaction has been measured at IUCF very precisely near the threshold \cite{pi0_iucf}. 
In theoretical calculations, it has been shown that the large contribution of the $s$-wave pion-production amplitude is necessary to reproduce the experimental data.
The proposed short-range effect between the nucleons gives an essential contribution into the $s$-wave amplitude where the final protons couple to a $S$-state \cite{Lee}. 
In order to understand the $s$-wave pion-production mechanisms systematical studies based on the chiral effective theory  are also still progressing \cite{chiral}.
Recently the investigations of the polarization observables have attracted interest from the viewpoint of the partial waves amplitude (PWA) analysis . 
These studies are expected to make a breakthrough in the elucidation of the origin of the large $s$-wave amplitude.
The analyzing power and spin correlation coefficients integrated over  the pion angle and energy have been obtained at four bombarding energies between 325 and 400 MeV \cite{pi0_iucf_pol}.
These data are compared with the theoretical calculations in \cite{hanhart2} and \cite{tamura}. 
Only these two currently existing models are able to predict polarization observables.
The calculations provide a good fit to the cross section close to the threshold but underestimate the contribution of the $s$-wave pion-production amplitude, where final protons are in a $P$-state, as deduced from the measurement of polarization observables. 
Therefore the origin of the $s$-wave pion-production amplitude is not clear for the $P$-state  as well as for the $S$-state. 

In this article, we report the experimental results of the analyzing power ($A_{y}$) as a function of the pion emission angle ($\theta_{\pi}$) in the center-of-mass system (C.M.S.) in order to obtain information about the relative strength of the s-wave pion production, where the final protons are in $P$-state and $S$-state.
Based on few partial wave amplitudes,
the deduced angular dependence of the $A_{y}$ $\times$ spin-averaged cross section is expressed in terms of the associated Legendre polynomials $P_{11}(\cos\theta_{\pi})$ and $P_{21}(\cos\theta_{\pi})$ which are symmetrical and asymmetrical around ${\theta}_{\pi}$ = 90$^{\circ}$, respectively.
The strength of the $P_{11}$ term corresponds to the contribution of the pion s-wave amplitude from a $P$-state, whereas that of the $P_{21}$ term is determined by the contribution from the $S$- and higher state.
The angular dependence is expected to be sensitive to the relative
strength between the $S$-state and $P$-state amplitudes and will give
useful information to make the origin of the $s$-wave amplitudes more clear.
The analyzing power is also shown as a function of the relative momentum of final protons with the angular integration, which enable one to select the term of the $P_{11}$ function and to observe the momentum dependence of the $P$-state amplitude.
The momentum dependence of the $P$-state amplitude is studied in terms of the interaction range of the pion production mechanism.

Experiment has been carried out using the 390 MeV polarized proton beam from Ring Cyclotron at Research Center for Nuclear Physics (RCNP), Osaka University, Japan.
The schematic view of the experimental setup is shown in Fig.1. 
The used liquid hydrogen target (L.H target) and cooling system has been developed by Kyushu University group \cite{target}.
The target thickness is 8.5 mm. 
The measurements with the gas hydrogen target have been also performed to study the amount of background events coming from construction materials.
The Faraday cup installed in the beam dump monitors the beam intensity.
The array of plastic scintillators is employed to detect the outgoing particles and measure the kinematical variables: scattering angles and energies of two protons, on the basis of coplanar geometries. 
The number of measured variables is sufficient to determine the kinematics of three-body final state. 
The energy of protons is measured by the plastic scintillator (E-counter) which can stop protons up to the 250 MeV kinetic energy.
The energy resolution of the E-counter was better than 2$\%$ at 200 MeV (FWHM).
The plastic scintillator hodoscope (Hodoscope) mounted in front of the E-counter is used to determine the direction of the outgoing particles. 
The angle covered by one hodoscope is $\pm$17 mrad horizontally and $\pm$30 mrad vertically. 
Our detector can measure outgoing protons in the scattering angular range of 15$^{\circ}-$35$^\circ$, which corresponds to the relative momentum of final protons from 150 MeV/c to the maximal kinematically allowed momentum ($\simeq$ 220 MeV/c) and covers the polar angle of the relative momentum from 40$^{\circ}$ to 140$^{\circ}$.
Anticoincidence counter (Anticounter) identifies the background events coming from the random coincidences of the $pp$ elastic scattering.
The beam polarization has been monitored by detecting the $pp$ elastic scattering event from the liquid hydrogen target using the set of scintillator counters (Polarimeter) placed at 60$^{\circ}\pm$1$^{\circ}$.
The analyzing power at the angle of 60$^{\circ}\pm$1$^{\circ}$ is found to be $-$0.36 from a database of SAID \cite{said}. 
The beam polarization during the experiment has been 65$-$75${\%}$. 

The ${\pi}^0$-production event is identified by the missing mass technique. 
The background due to the random coincidence of the  elastic scattering events and  inelastic scattering events from other construction materials has been subtracted.  
After subtracting background, the events that do not deposit the full energy on the E-counter still are left as a tail in the missing mass spectrum. 
These events are caused by the elastic scattering and charge exchange nuclear reaction in the scintillator (nonfull-peak event).
The tail is involved into the systematical error. 

Figure 2 shows the angular dependence of $A_{y}$ for three relative-momentum regions.
The errors in the figure are only systematical mainly coming from the energy measurement of the E-counters and the estimation of the nonfull-peak events.
The statistical error is below 5$\%$.
In order to obtain the relative strength of the coefficients of the Legendre polynomials $P_{11}$ and $P_{21}$, the experimental data has been fitted by
\begin{eqnarray}
~A_{y} =  [{a} ~P_{11}(\cos\theta_{\pi})  
         + {b} ~P_{21}(\cos\theta_{\pi}) ]
/{\sigma}(\theta_{\pi}), 
\label{eq:ay}
\end{eqnarray}
where $a$ and $b$ are the strength parameters.
 The spin-averaged angular distribution $\sigma({\theta_{\pi}})$ is obtained independently by fitting the experimental data on the spin-averaged angular distribution with the function of $\sigma_0 (1+ {c} \cos^2\theta_{\pi})$. 
The $\sigma_0$ and $c$ are free parameters, and the magnitude of 
the $c$ is 0.5-0.6. 
In Fig.2, dashed lines show the results of the fitting with Eq.(\ref{eq:ay}).
The main contribution comes from the $P_{11}$ term and the fraction of the $P_{21}$ term to the contribution of the $P_{11}$ is below 20$\%$, which makes the angular dependence of $A_{y}$ slightly asymmetrical.
IUCF data at close energies also show a large $P_{11}$ component \cite{pi0_iucf_new}.
In terms of few PWAs ($\bf Ss$,$\bf Ps$,$\bf Pp$,$\bf Sd$), 
the coefficient $a$ is determined by the $\bf Ps \times \bf Pp$ 
whereas the coefficient $b$ is determined by the $\bf Ss \times \bf Sd$ and $\bf |Pp|^2$ amplitudes. 
Here, the capital and small letters show the angular momentum state of final protons and pion, respectively. 
Therefore the main contribution of the $P_{11}$ term indicates that the strength of $P$-state amplitude dominates over the $S$-state amplitude. 
On the other hand the model calculations of RCNP group (Fig.2) show more asymmetrical behavior than the experimental results and the model calculations overestimates the data.
That means the calculated strength of the $P$-state amplitude is much smaller than the experimental one.
More theoretical studies are needed for elucidating the origin of this disagreement.

Figures 3 (a) and (b) show the relative momentum distribution (dN/dk) and integrated analyzing power as a function of the relative momentum of final protons ($k$), respectively.
The value of $A_{y}$ increases with $k$.
Since the term of the $P_{21}$ function in Eq.(\ref{eq:ay}) becomes zero after the integration over the pion angle, the dependence of the $A_{y}$ can be expressed by the $P$-state amplitudes and  dN/dk as 
\begin{eqnarray}
A_{y} = {\bf Ps \times Pp} ~\rho ({k}) /({dN / d k}) ,
\label{eq:aymon}
\end{eqnarray}
where $\rho ({k})$ is a phase space factor. Thus 
one can find from Eq.(\ref{eq:aymon}) that the momentum dependence of $P$-state amplitude can be obtained utilizing the known dependence of dN/dk and $\rho ({k})$.
From the theoretical point of view,
the PWA is calculated by $\int dr r^2 u_{f}^{k}(r) \phi_{l_{\pi}}^{q}(r) {\hat \Pi}(r) u_{i}^{p}(r)$, i.e. the overlap integral between the pion-production operators ${\hat \Pi}(r)$ and the radial wave function of initial protons $u_{i}^{p}(r)$, final protons $u_{f}^{k}(r)$, and pion  $\phi_{l_{\pi}}^{q}(r)$, where $p$ and $q$ is the momentum of initial protons and of a pion in C.M.S, respectively. ${l_{\pi}}$ is the angular momentum of a pion.
For example, the analytic forms of the $s$-wave pion-production amplitude where the final protons are in $S$-state are presented in Ref.\cite{horowitz}.
According to the argument of Ref.\cite{pi0_iucf_new},
the strong momentum dependence of the $P$-state amplitude comes from the dependence of the radial wave function of final protons and a pion, and the pion-production operator is commonly considered as momentum independent.
In the overlap integral the pion-production operator selects typical region of the radial wave functions when integrated over the relative distance of protons.
Accordingly, the radial wave function of final state shows different momentum dependence on that selected region.
Therefore the wave function of final state with a fixed distance of protons gives a general estimation about the typical interaction region of the pion-production operator in a certain PWA.  
For this purpose,   
the dependence of the $P$-state amplitude with a fixed distance ($r$) between protons is estimated by 
${\bf Ps \times Pp} \sim (j_{1}(kr))^{2} j_{1}({1\over 2}qr) j_{0}({1\over 2}qr)$, where $j_{l}(x)$ is  the spherical Bessel functions with an orbital angular momentum  ($l$), which is used for the undistorted wave function of final protons and pion 
(distorted wave function for $P$-state protons does not change the final result).
Results are shown in Fig.3 (b).
The value of dN/dk is taken from the solid line of Fig.3 (a) that gives the best fit to the data. 
The calculations performed at several distances (r=1,2,3,3.5 fm) show that the momentum dependence becomes close to the experimental data as the distance increases. 
The r$\sim$3 fm  (dash-double-dotted line) is more preferable.
The long range part of the wave function gives the proper dependence and it implies that the contribution of the long range production mechanism is more preferable than the short range one r$\sim$1 fm (dotted line).
One must keep in mind that the above discussions are based an assumption that the momentum dependence of the pion-production operator is quite small.
Therefore, the long range mechanism might not be the only way to explain the momentum dependence of experimental data.
However as suggested by Ref.\cite{tamura}, the discrepancy between the theoretical calculations and experimental data on the analyzing power presented in this article (both  angular dependence and momentum dependence in Fig.3 (b)) indicates the necessity of some long-range mechanisms for $s$-wave pion production in $P$-state amplitude to improve the predicted $A_y$.   
Such a behavior corresponds to the expectation since the probability of the $P$-state at short distance is much smaller than the probability of $S$-state and therefore the short-range mechanism does not support the $P$-state amplitude. 

In summary, 
we have measured the angular and momentum dependence of the analyzing power for the ${\vec p}p\rightarrow pp{\pi}^0$ reaction at the incident energy of 390 MeV . 
The angular dependence shows the dominated contribution of $P_{11}$ and that the contribution of the $s$-wave pion production comes mainly from the $P$-state of final protons at this energy.
The long range part of the $P$-state wave function gives a general explanation to the experimental momentum behavior of analyzing power  and it is deduced that the long range interaction is important for the pion production from the $P$-state nucleons and 
further study is needed to pin down the production mechanism with a long range interaction.  

The authors are grateful to the RCNP cyclotron staff for 
their support throughout the presented experiments.
We acknowledge K. Sagara for the liquid hydrogen
target system.
This experiment was performed under the programmed No. E140
at RCNP.

\newpage

Fig. 1. Experimental setup. The shadow region shows liquid hydrogen target.

Fig. 2. The angular dependence of the analyzing power of the ${\vec p}p\rightarrow pp{\pi}^0$ reaction for three regions of relative  momentum of final protons at the beam energy of 390 MeV.
The momenta range is shown at the top of figures. 
The horizontal axis shows the emitted angle of a pion in C.M.S. 
Solid lines show the result of RCNP model  and
shaded area indicates uncertainties 
for the calculation \cite{tamura}.  
Dashed
lines show the  fitted result of the Legendre polynomial, see Eq.(\ref{eq:ay}).

Fig. 3. (a) Relative momentum distribution of final protons.
The vertical axis shows the normalized yields  in arbitrary unit.
The solid line is the fitted result.
(b) Relative momentum dependence of the analyzing power.
The solid line shows the model calculations (same one of Fig.2).
The dotted, dash-dotted, dashed-double-dotted and dashed lines 
show the results with Bessel function at distances $r$=1, 2, 3 and 3.5 fm, respectively.

\newpage

\begin{figure}[h]
\begin{center}
\epsfig{file=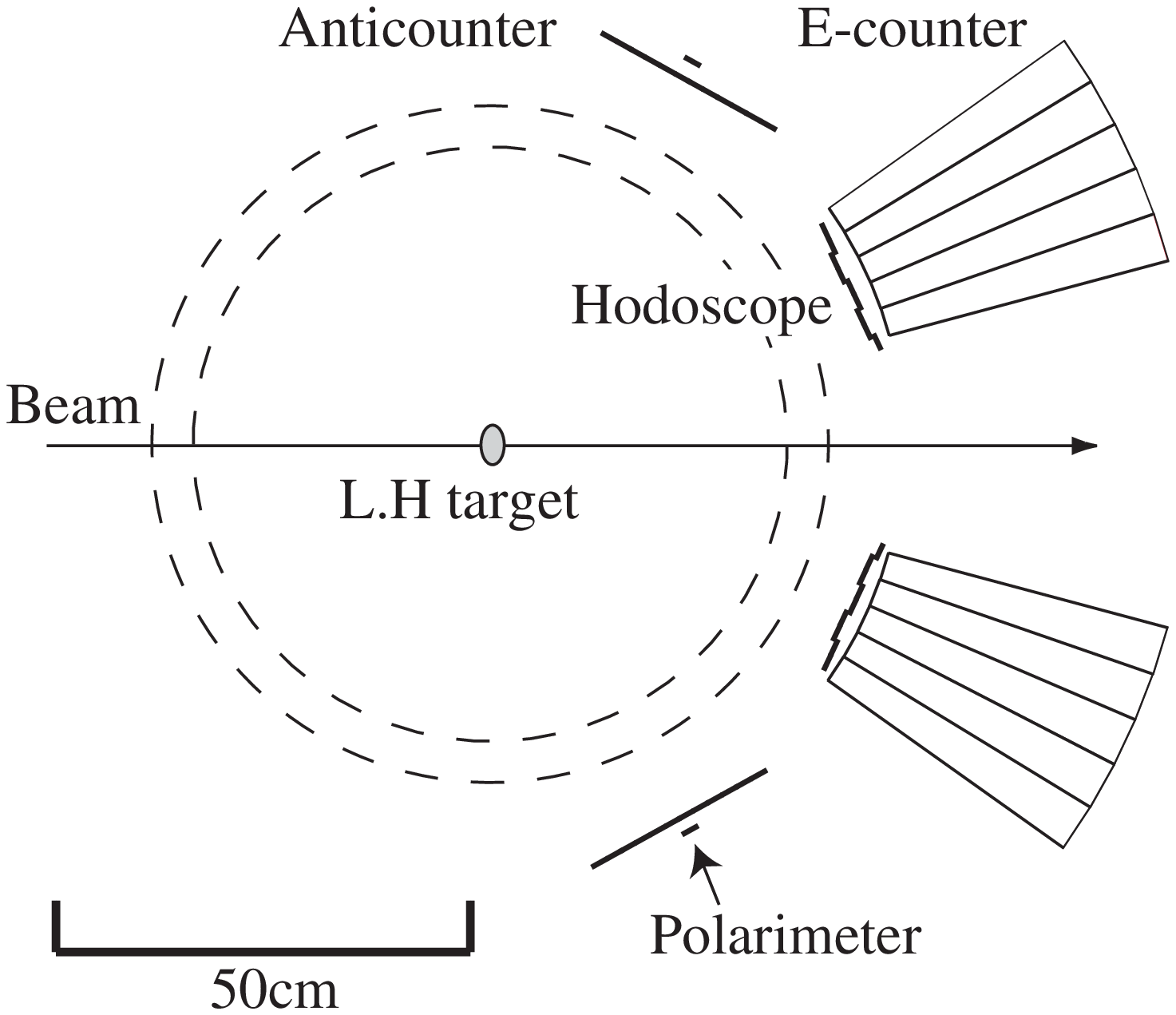,width=12cm}
\end{center}
{Fig.1}
\end{figure}

\newpage

\begin{figure}[h]
\begin{center}
\epsfig{file=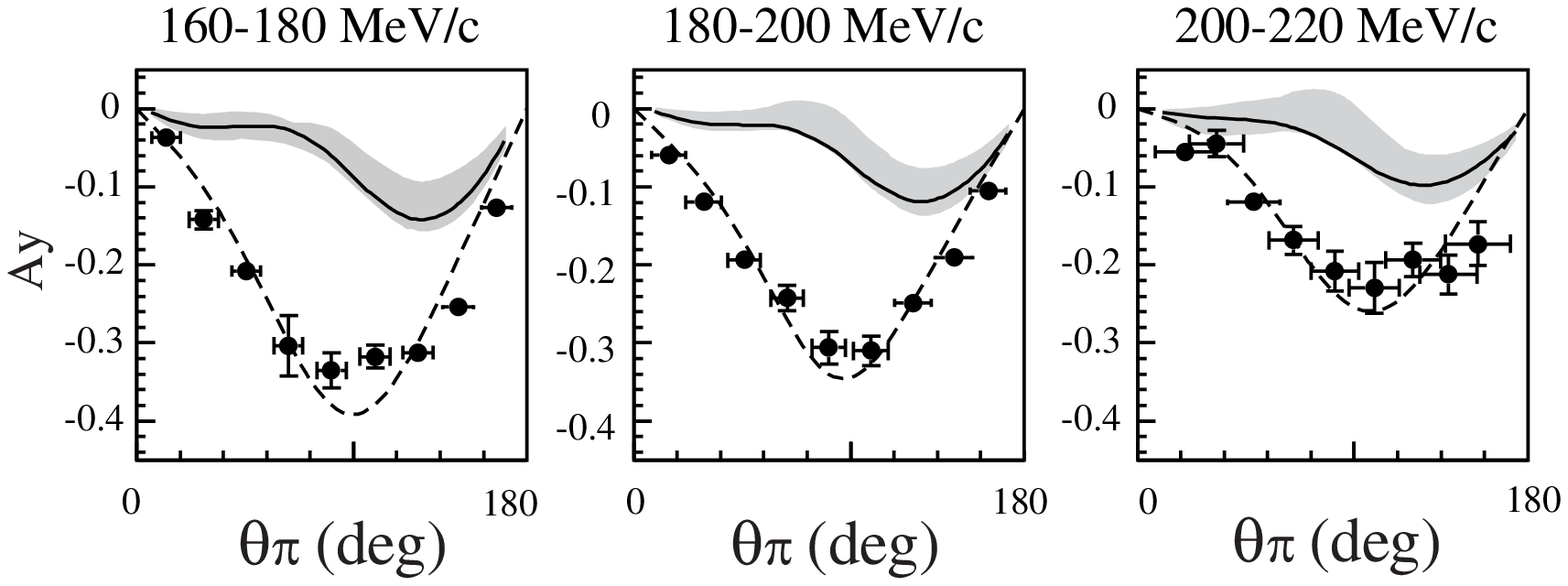,width=15cm}
\end{center}
{Fig.2}
\end{figure}

\newpage
\begin{figure}[h]
\begin{center}
\epsfig{file=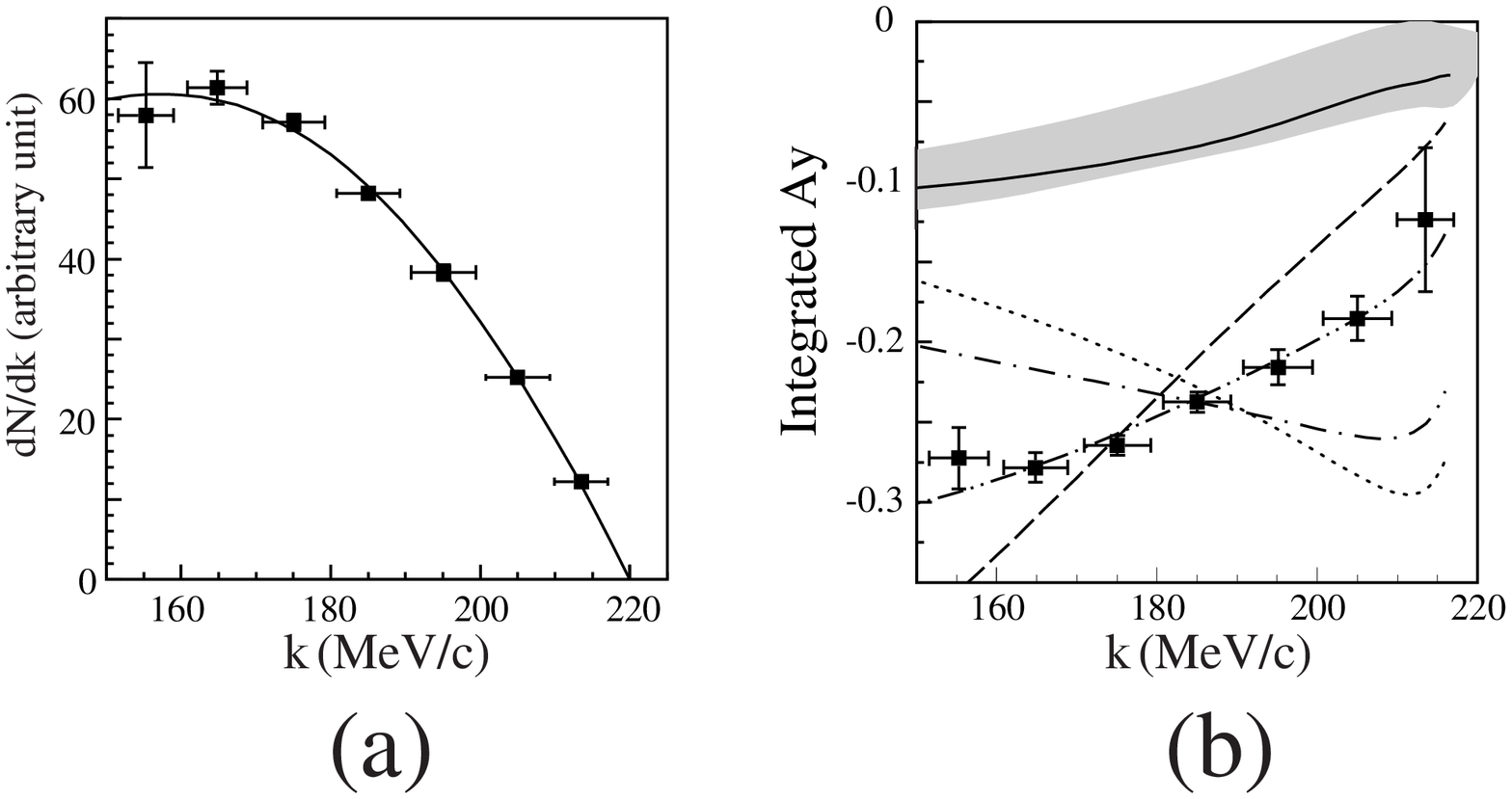,width=15cm}
\end{center}
{Fig.3}
\end{figure}

\end{document}